\documentclass[ALICE,manyauthors]{cernphprep}
\RequirePackage{ifpdf}                                                                                                                                                                                     
\pdfoutput=1
\usepackage{graphicx}
\usepackage{amsmath}
\usepackage{amssymb}
\usepackage{xspace}
\usepackage{float}
\usepackage{placeins}
\usepackage{lineno}
\usepackage{etoolbox}
\usepackage{version}
\usepackage[pdftex,usenames,dvipsnames]{xcolor}

\ifpdf
  \usepackage[pdftex]{hyperref}
  \hypersetup{%
  colorlinks = true,
  linkcolor  = black
}
  \hypersetup{
    final,
    a4paper=true,
    colorlinks=true,
    bookmarks=true,
    pdffitwindow=true,
    pdfnewwindow=true,	
    pdfauthor={ALICE Collaboration},
    pdfcreator={pdflatex},
    pdfproducer={ALICE Collaboration},
    pdftitle={Determination of the event collision time in ALICE at the LHC},
    pdfsubject={Determination of the event collision time in ALICE at the LHC},
    pdfkeywords={}
  }

\else
  \usepackage[pdftex]{hyperref}
\fi

\newcommand {\sqrtsNN}   {\ensuremath{\sqrt{s_{\textsc{NN}}}}\xspace}
\newcommand {\sqrts}     {\ensuremath{\sqrt{s}}\xspace}
\newcommand {\pp}        {\mbox{$\mathrm {p\kern-0.05em p}$}\xspace}
\newcommand {\PbPb}      {\ensuremath{\mbox{Pb--Pb}}\xspace}
\newcommand {\pPb}       {\ensuremath{\mbox{p--Pb}}\xspace}
\newcommand {\gmom}     {\mbox{\rm GeV$\kern-0.15em /\kern-0.12em c$}}

\newcommand{\TeV}{\ensuremath{\mathrm{TeV}}\xspace}
\newcommand{\gevc}{\ensuremath{\mathrm{GeV}/c}\xspace}

\newcommand{\run}[1]{\textsc{Run\,#1}}

\newcommand{\tev}{$t_{\rm{ev}}$\xspace}
\newcommand{\avetev}{$\langle t_{\rm{ev}}\rangle$\xspace}
\newcommand{\tcoll}{$t_{\rm{ev}}$\xspace}
\newcommand{\TOTO}{$t_{\rm{ev}}^{\rm{T0}}$\xspace}
\newcommand{\TOTOF}{$t_{\rm{ev}}^{\rm{TOF}}$\xspace}
\newcommand{\TOBest}{$t_{\rm{ev}}^{\rm{Best}}$\xspace}
\newcommand{\TOFILL}{$t_{\rm{ev}}^{\rm{Fill}}$\xspace}
\newcommand{\TOTOAC}{$t_{\rm{ev}}^{\rm{T0AC}}$\xspace}
\newcommand{\TOTOA}{$t_{\rm{ev}}^{\rm{T0A}}$\xspace}
\newcommand{\TOTOC}{$t_{\rm{ev}}^{\rm{T0C}}$\xspace}

\newcommand{\sigmatcoll}{$\sigma_{t_{\rm{ev}}}$\xspace}
\newcommand{\sigmaTOTO}{$\sigma_{t_{\rm{ev}}^{\rm{T0}}}$\xspace}
\newcommand{\sigmaTOTOA}{$\sigma_{t_{\rm{ev}}^{\rm{T0A}}}$\xspace}
\newcommand{\sigmaTOTOC}{$\sigma_{t_{\rm{ev}}^{\rm{T0C}}}$\xspace}
\newcommand{\sigmaTOTOAC}{$\sigma_{t_{\rm{ev}}^{\rm{T0AC}}}$\xspace}

\newcommand{\sigmaTOTOF}{$\sigma_{t_{\rm{ev}}^{\rm{TOF}}}$\xspace}
\newcommand{\sigmaTOBest}{$\sigma_{t_{\rm{ev}}^{\rm{Best}}}$\xspace}
\newcommand{\sigmaTOFILL}{$\sigma_{t_{\rm{ev}}^{\rm{Fill}}}$\xspace}

\newcommand{\tTOF}{$t_{\rm{TOF}}$\xspace}
\newcommand{\tEXP}{$t_{\rm{exp},i}$\xspace}

\newcommand{\tof}{time-of-flight\xspace}
\newcommand{\TOF}{Time-of-Flight\xspace}

\begin{document}
\begin{titlepage}
\PHyear{2016}
\PHnumber{253}
\PHdate{27 September}

%\linenumbers 
\title{Determination of the event collision time\\with the ALICE detector at the LHC}

\Collaboration{ALICE Collaboration\thanks{See Appendix~\ref{app:collab} for the list of collaboration members}}
\ShortAuthor{ALICE Collaboration} % appears on left page headers, do not change
\begin{abstract}
Particle identification is an important feature of the ALICE detector at the LHC.
In particular, for particle identification via the \tof technique, the precise 
determination of the event collision time represents an important
ingredient of the quality of the measurement.
In this paper, the different methods used for such a measurement in ALICE by means of the T0 and
the TOF detectors are reviewed. Efficiencies, resolution and the improvement of the
particle identification separation power of
the methods used  are presented for the different LHC colliding
systems (\pp , \pPb and \PbPb) during the first period of data taking of
LHC (\run{1}).
\end{abstract}
\end{titlepage}
\setcounter{page}{2}

\maketitle
\section{Introduction}
\label{sec:intro}
The main task of the ALICE experiment~\cite{ppr1,ppr2}  at the LHC is the
 study of the properties of
the strongly interacting, dense and hot matter created in high-energy heavy-ion collisions.
Many physics analyses are based on the
capability of the ALICE detector to perform Particle IDentification (PID) using different and complementary techniques.
In the intermediate momentum range (from 0.5 to 3-4 \gmom) this task is
mainly accomplished using the \tof measurements which rely on a precise
determination of the event collision time, the track
length and momentum, and the arrival time of the tracks to the \TOF (TOF) detector. 

The track length and momentum measurement is defined by the Inner Tracking System (ITS) and the Time Projection Chamber (TPC)~\cite{aliceperf}. The ITS is composed of six cylindrical layers of silicon detectors, located at radial distances between 3.9 and 43 cm from the beam axis. The TPC is a large volume cylindrical chamber with high-granularity readout that surrounds the ITS covering the region 85 $< r <$ 247 cm and --250 $< z <$ 250  cm in the radial $r$ and longitudinal $z$ directions, respectively.  These detectors, covering the pseudo-rapidity interval --$0.9 \le \eta \le 0.9$ for tracks reaching the outer layer of the TPC, also provide PID information via the specific energy loss (d$E$/d$x$) measurements.

The measurement of the \tof of the tracks is based on the TOF detector. On the other hand, the event collision time \tev is determined with the information coming from both the TOF and the T0 detectors. 

\begin{figure}
\centering
\includegraphics[width=10cm]{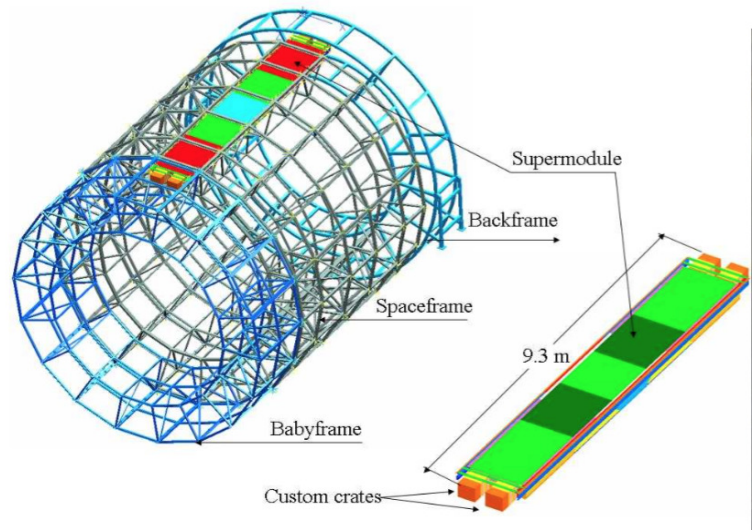}
\caption{A schematic layout of one of the 18 TOF supermodules inside the ALICE spaceframe.}
\label{fig:TOFlayout}
\end{figure}

The TOF system~\cite{tofprod}  covers the pseudo-rapidity interval --$0.9 \le \eta \le 0.9$ and full azimuthal acceptance. The system is located, according to a cylindrical symmetry, at an average distance of 3.8 m from the beam pipe spanning an active area of 141 m$^2$.
The detector is made of 1593 Multi-gap Resistive Plate Chambers (MRPC), with a sensitive area of 7.4$\times$120 cm$^2$ each.  Each MRPC is segmented into 96 readout pads of area
2.5$\times$3.5 cm$^2$. The MRPCs are packed then in five modules for each of the 18
azimuthal sectors of the ALICE spaceframe in a ``TOF supermodule'', as shown in Fig.~\ref{fig:TOFlayout}.
This detector has a time resolution of $\sim$80 ps during the data taking~\cite{tofperf}. 

\begin{figure}
\centering
\includegraphics[width=10cm]{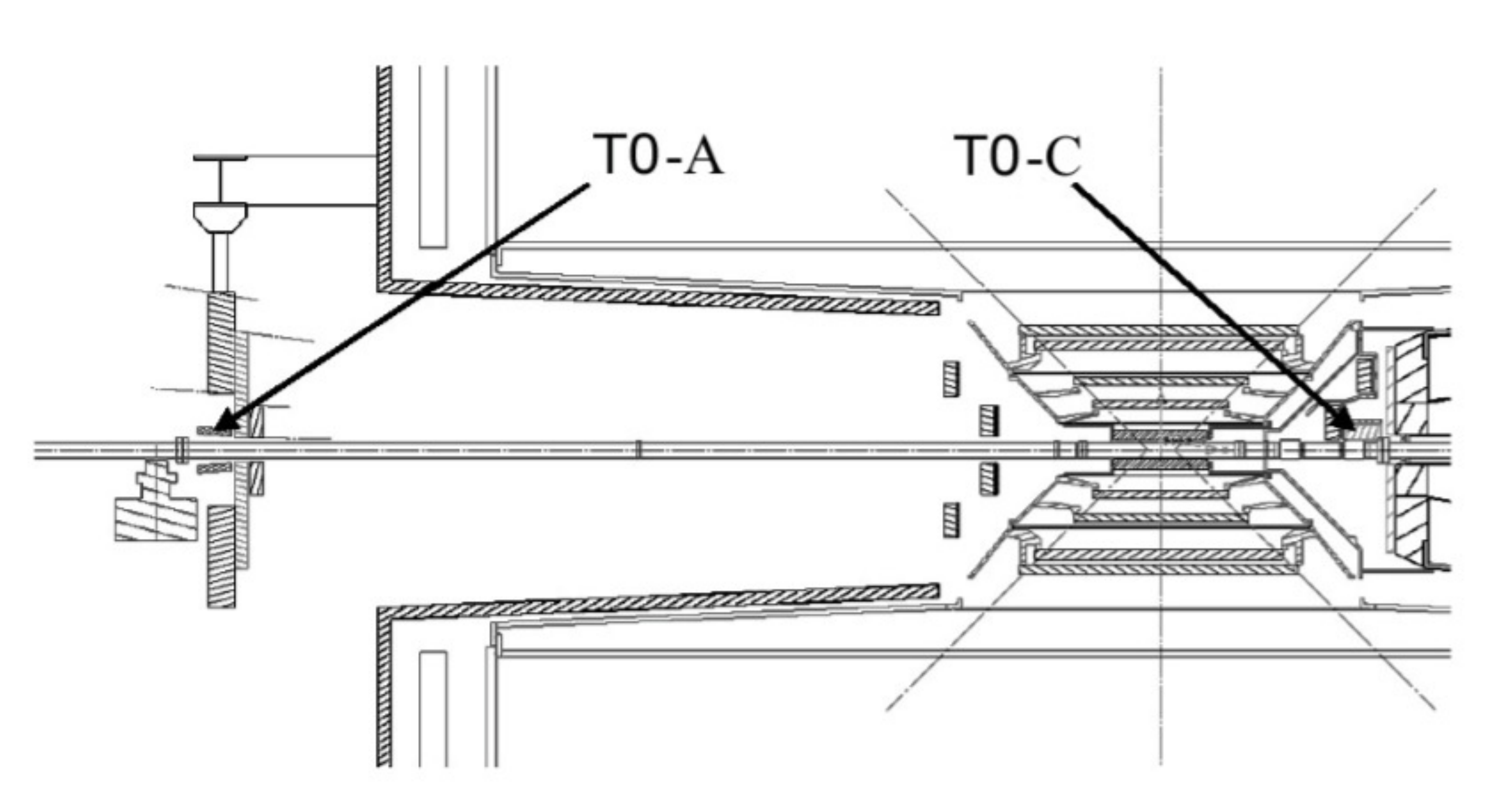}
\caption{The layout of the T0 detector arrays inside ALICE.}
\label{fig:T0layout}
\end{figure}

The T0 detector~\cite{t0dect} consists of two arrays of
Cherenkov counters T0A and T0C, positioned on both sides of the interaction point (IP)  at a distance of 374 cm and --70 cm (as shown in Fig.~\ref{fig:T0layout}), covering the pseudorapidity region 4.61 $ \le \eta \le $ 4.92 and --3.28 $ \le \eta \le $ --2.97, respectively. The small distance from the IP for T0C had to be chosen because of the space constraints imposed by the front cone of the muon absorber and other forward detectors. On the opposite side the distance of the array T0-A is comfortably far from the congested central region. 

Each array has 12 cylindrical counters equipped with a quartz radiator 20 mm in diameter and 20 mm thick and a photomultiplier tube. The T0 detector provides a measurement of the \tev. It also provides the collision trigger and 
monitors the luminosity providing fast feedback to the LHC accelerator team. The measured time resolution of the T0 detector is $\sim$50 ps for single MIP events and reaches $\sim$25 ps at higher multiplicities.

The TOF and the T0 detectors use different front-end electronics but the same digital electronics. The latter is based on the HPTDC (High Performance Time Digital Converter)~\cite{hptdc} developed by the  CERN Microelectronic Group for LHC experiments. The time measurement is performed with 25 ps bin width resolution with respect to the trigger time, latched with the 40 MHz LHC clock phase. The measurement corresponds for this application to an ionizing particle hit in the TOF MRPC or a photon hit in the T0 photomultipliers. The HPTDC is free running and hit time measurements are stored
in internal buffers within a given latency window, waiting for the trigger
arrival.

Relevant for the following discussion is also the V0 detector. It consists of two scintillator arrays built around
the beam pipe covering the pseudorapidity ranges 2.8$ \le \eta \le $5.1 (V0A) and --3.7$ \le \eta \le $1.7 (V0C) is used for triggering and event selection. In \pPb collisions it is also used to define the multiplicity of the collision exploiting  the information from the amplitude of the signal measured by the V0A scintillators \cite{Adam:2014qja} while in \PbPb it is used to define the centrality through the summed amplitudes in the V0 scintillators as described in~\cite{Abelev:2013qoq}.

The particle identification with the TOF detector is based on the
comparison between the \tof of the track from the primary vertex to the
TOF detector and the
expected time under a given mass hypothesis \tEXP (i = e, $\mu$,
$\pi$, K, p, d, t, $^{3}$He, $^{4}$He). The former is
defined as the difference between the arrival time \tTOF
measured by the TOF detector itself and the event collision time \tcoll. The expected time is the
time it would take for a particle of mass $m_{i}$ to go from the
interaction point to the TOF. To take into account the energy loss and
the consequent variation in the track momentum, \tEXP is calculated as
the sum of the small time increments $\Delta t_{i,k}$, each of which is
the time a particle of mass $m_{i}$ and momentum $p_{k}$ spends to
travel along each propagation step $k$ of lenght $\Delta l_{k}$
during the track reconstruction procedure:
\begin{equation}
  t_{\rm{exp},i}=\sum_{k}\Delta
  t_{i,k}=\sum_{k}\frac{\sqrt{p_{k}^{2}+m_{i}^{2}}}{p_{k}}\Delta l
  _{k} .
\end{equation}  

Therefore, the fundamental variable for the TOF PID is
$t_{\rm{TOF}}-t_{\rm{ev}}-t_{\rm{exp},i}$. Its resolution is
\begin{equation}
\label{eq:sigmapid}
\sigma_{\rm{PID},i}^{2}=\sigma^{2}_{t_{\rm{TOF}}}+\sigma^{2}_{t_{\rm{ev}}}+\sigma^{2}_{t_{\rm{exp},i}} .
\end{equation}

As mentioned earlier, the TOF detector resolution ($\sigma_{t_{\rm{TOF}}}$) is
$\sim$80 ps while the
uncertainty ($\sigma_{t_{\rm{exp},i}}$) due to the tracking and reconstruction, that includes estimates
of the energy losses through the material, depends on the momentum and on the particle species~\cite{tofperf}. The uncertainty on the event collision time (\sigmatcoll) depends on the method used to
determine it in the given event. 

The simplest PID estimator for a given mass hypothesis $m_{i}$ is then constructed as an $n\sigma$ quantity in the following way:
\begin{equation}
\label{eq:pidres}
n\sigma_{\rm{TOF},i}=\frac{t_{\rm{TOF}}-t_{\rm{ev}}-t_{\rm{exp},i}}{\sigma_{\rm{PID},i}} .
\end{equation}

This paper focuses on a fundamental term for the TOF PID determination: the event collision time \tev. The methods used for its determination are described in detail in the following sections. Their resolutions, efficiencies and impacts on the PID performance are reported for data samples collected in the different collision systems during \run{1}: \pp data at a center-of-mass energy of \sqrts~=~7~\TeV, \pPb data at a center-of-mass energy per nucleon pair of \sqrtsNN~=~5.02~\TeV and \PbPb at \sqrtsNN~=~2.76~\TeV. 
In Sec.~\ref{sec:datasample} the event and track selection are described, in Sec.~\ref{sec:avetevCal} the calibration and timing alignment procedure of the TOF with respect to the LHC clock and in Sec.~\ref{sec:methods} the methods for the determination of the event collision time \tev. Finally, in Sec.~\ref{sec:results} results for efficiencies, resolutions and impact on PID
separation power are presented and discussed. More informations on the general performances of the ALICE detectors in the first period of data taking at LHC are available in~\cite{aliceperf} . 

\section{Event and track selection}
\label{sec:datasample}
For the study reported in this paper the data were selected using a minimum bias trigger  based on the V0 detector. Events are further required to have a primary vertex reconstructed either from the tracks reconstructed both in the ITS and in the TPC or from the tracklets, which are track segments built from pairs of hits in the two innermost layers of the ITS. Only events with a reconstructed primary vertex within 10 cm from the nominal interaction point along the beam directions were used in the analysis. Furthermore, events with multiple reconstructed vertices were rejected, leading
to a negligible amount of pile-up events for all the colliding systems \cite{aliceperf}.
Finally, since the event collision time is a measurement that is needed to identify particles by means of the \tof technique performed by the TOF detector, only events with at least one track associated with a hit in the TOF detector are selected.
The number of analyzed events after these cuts is 12 millions for \pp at \sqrts~=~7~TeV, 10 millions for \pPb  and 1 million for \PbPb that are only a subsample of the available statistics collected by ALICE.

The performance of the event collision time will be reported in terms of the TOF track multiplicity of the event, that is the number of tracks associated with a hit on the TOF detector. This choice is driven by the fact that a hit on the TOF is the minimal request that a track has to satisfy to be identified via the \tof procedure.
For \PbPb events, the \tev measurement performance is also reported in terms of centrality, determined by the sum of the V0 amplitudes and
defined in terms of percentiles of the total hadronic \PbPb cross section~\cite{Abelev:2013qoq}, while for \pPb in terms of the V0A multiplicity~\cite{Adam:2014qja}.

\begin{figure}[t!]
\centering
\includegraphics[width=10cm]{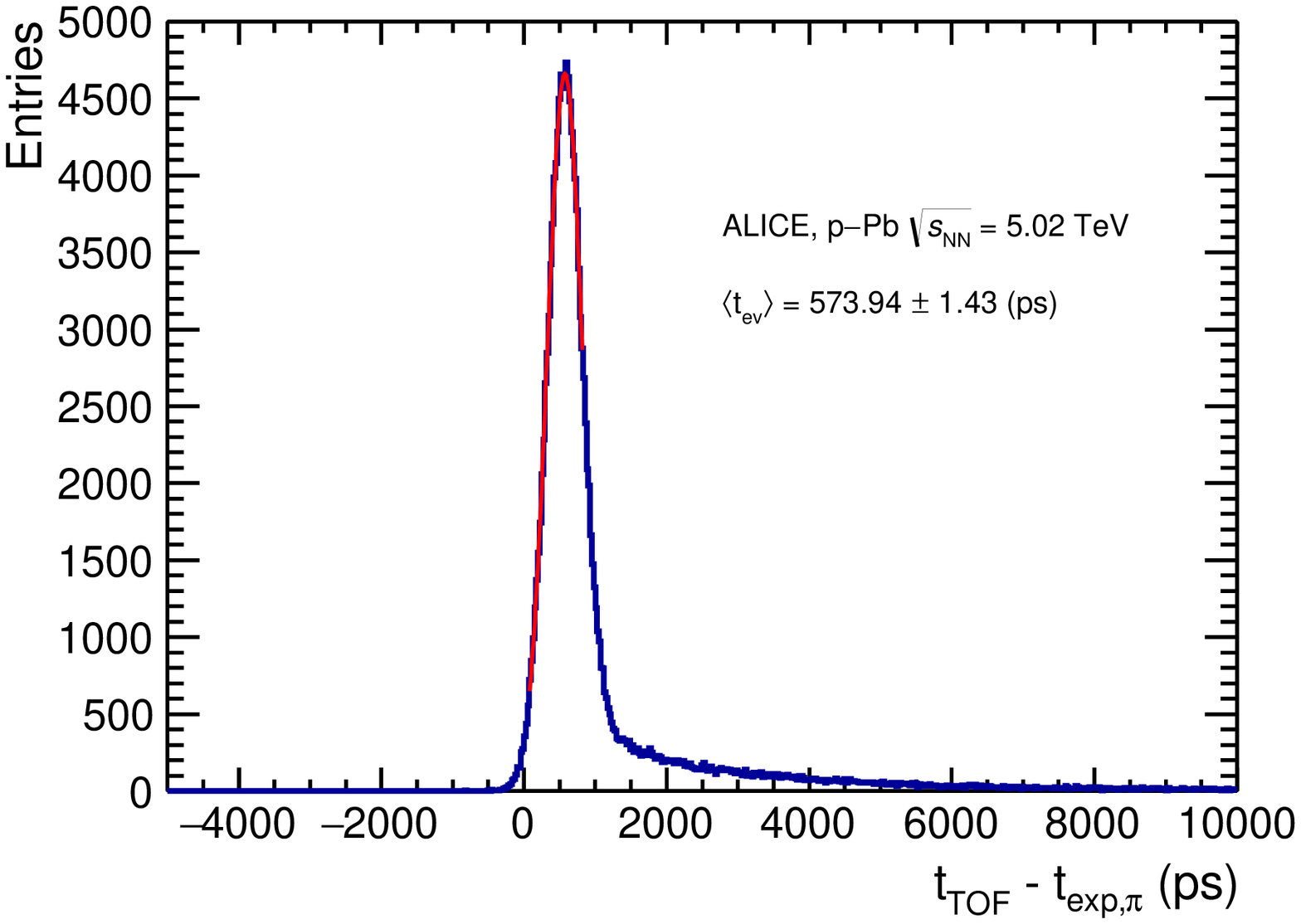}
\caption{Average collision time \avetev calculated for five minutes of \pPb data taken at \sqrtsNN $= 5.02$ \TeV.}
\label{fig:T0Fill}
\end{figure}

\section{TOF time alignment and calibration}
\label{sec:avetevCal}
As described in~\cite{tofperf}, the TOF signals are first calibrated for the
channel--by--channel offsets (which take into account the differences
due to the cable length) and the time--slewing effects.
Then, to align the \tof with respect to the
LHC clock, a global shift with respect to the clock phase, \avetev, is
calculated by the TOF itself, for each LHC fill, during the calibration procedure as
described below and applied  as a global offset to all the measured times.

Due to the fact that the phase of the LHC clock during a fill,
as distributed to the experiments, is subject to shifts correlated with the environment temperature (the refractive index of the fibers used for the clock distribution has a dependency on the temperature), \avetev is calculated with a five minutes granularity in time. This interval is increased in steps of five minutes if the number of events in the interval is smaller than 1000 or the number of tracks selected for the procedure is smaller than 20000.
The \tof measured for the selected tracks
is then compared to the \tEXP obtained assuming the pion mass
hypothesis. The choice of using the pion mass as reference is justified by the fact that pions are the most abundant species produced in the collisions, and they largely dominate the time spectrum distribution. The difference between the measured \tof and the expected times is fitted with a Gaussian function. Its mean
corresponds to the global offset to be applied to all the \tof signals
measured in the time interval under study, in order to align the \tTOF with respect to the
LHC clock. Figure~\ref{fig:T0Fill} shows
an example of such a fit for \pPb data at \sqrtsNN $= 5.02$ \TeV
collected in 2013.  

\begin{figure}[t!]
\centering
\includegraphics[width=10cm]{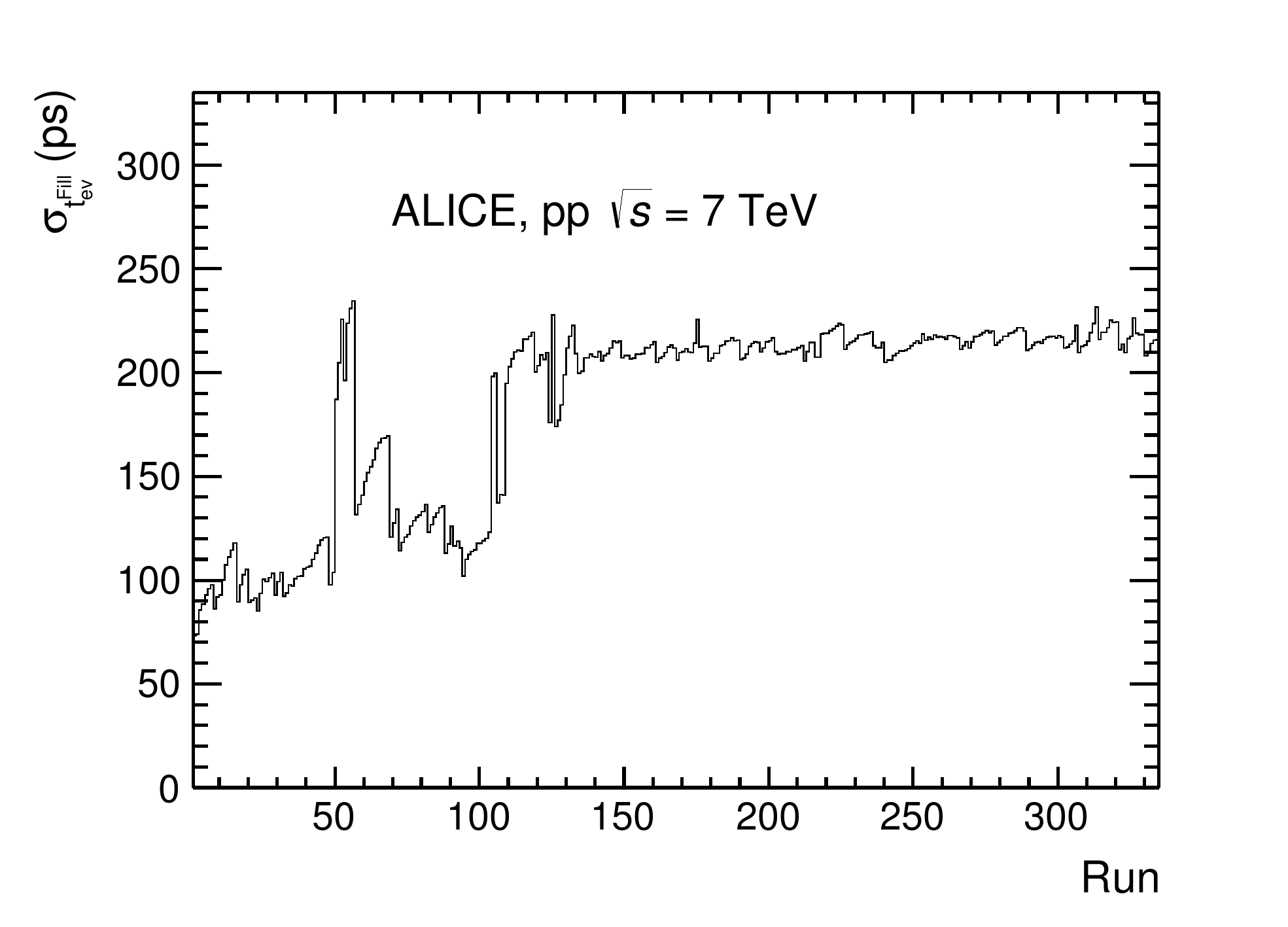}
\caption{Resolution of the \TOFILL (\sigmaTOFILL) for all 335 \pp runs recorded at \sqrts~=~7~TeV in 2010.}
\label{fig:sigmaT0Fill}
\end{figure}

\section{Methods for the event-by-event collision time determination}
\label{sec:methods}
Since the bunches have a small but finite size and it is not known which
of the particles in the bunches have collided, the event collision
time has a natural spread with respect to the nominal beam
crossing. Therefore, an event
time \tev has to be measured on an event-by-event basis. 
If the event-by-event procedures described below can not be used, \tev 
is set to zero. Conventionally, this null value is named \TOFILL. It is assumed null because \avetev has been
already subtracted as part of the calibration procedure described in Sec.~\ref{sec:avetevCal}.
Its resolution is directly connected to the vertex
spread along the beam direction estimated by the ITS per run and derived via \sigmaTOFILL= $\sigma_{\rm{vertex}}/\kern-0.12em c$.
In Figure~\ref{fig:sigmaT0Fill} the \sigmaTOFILL is reported for all the
runs of \pp collisions at \sqrts~=~7~\TeV collected during the 2010 data taking.

The variation of \sigmaTOFILL shown in Fig.~\ref{fig:sigmaT0Fill} depends 
on the beam optic configurations. After the initial LHC operations \sigmaTOFILL became more or less constant at $\sim$200
ps. Therefore, if \tev can not be computed on an event-by-event basis,
\tev is set to \TOFILL which has a resolution of  $\sim$200 ps. This
becomes then the dominant term in the TOF PID resolution (see Eq.~\ref{eq:sigmapid}).

To improve the TOF PID performance on an event-by-event basis reducing the \sigmatcoll in
Eq.~\ref{eq:sigmapid} with respect to the value of \sigmaTOFILL, the
\tev can be computed by the TOF itself
(\TOTOF), by the T0 detector (\TOTO) or by a combination of the two
(\TOBest) as shown in the following sections.

\subsection{Event collision time measurement performed by the TOF detector}
\label{sec:TOFmethods}
The event collision time is estimated  by the TOF detector (\TOTOF) on an event-by-event
basis by means of a $\chi^{2}$-minimization procedure. Having in the
event $n_{\rm{tracks}}$ matched to a corresponding hit on the TOF detector and satisfying
basic quality cuts, it is possible to define certain combinations of
masses $\vec{m_{i}}$ assigning independently for each track the $\pi$,
K or p mass. The index $i$ indicates one of the possible combination $(m_{1},
m_{2}, ..., m_{n_{\rm{tracks}}})$ among the $3^{n_{\rm{tracks}}}$ones.

For each track the following weight is evaluated
\begin{equation}
\label{eq:wj}  W_{i}=\frac{1}{\sigma_{\rm TOF}^2+\sigma_{t_{\rm {exp},i}}^2} .
 \end{equation}

The event time is then deduced as in Eq.~\ref{eq:tev} where the track
index is omitted for simplicity
\begin{equation}
\label{eq:tev} 
t_{\rm{ev}}^{\rm{TOF}}(\vec{m_{i}})= \frac{ \sum\limits_{n_{\rm{tracks}}}W_i (t_{\rm TOF} -t_{\rm {exp},i})} 
 { \sum\limits_{n_{\rm{tracks}}}W_i} 
\end{equation}
and the resolution is given by 
\begin{equation}  
\label{eq:sigTOF} 
 \sigma_{t_{\rm{ev}}^{\rm{TOF}}}(\vec{m_{i}}) = \sqrt{\frac{1}{  \sum\limits_{n_{\rm{tracks}}}W_i}}.
 \end{equation}

The  following $\rm{\chi^2}$ is then calculated

\begin{equation}
\label{eq:t0-TOF} \chi^{2}(\vec{m_{i}})=\sum_{n_{\rm{tracks}}}\frac{ ((t_{\rm
TOF}- t_{\rm{ev}}^{\rm{TOF}}(\vec{m_{i}})) -t_{\rm {exp},i})^2}{\sigma_{\rm {TOF}}^2+
\sigma_{t_{\rm {exp},i}}^2} .
 \end{equation}

The combination $\vec{m_{i}}$ that minimizes this $\chi^2$ is 
used to derive \TOTOF via Eq.~\ref{eq:tev}. 

This general procedure is refined in two ways.
To avoid possible PID biases which are important especially in low
multiplicity events, a track can not be used to compute the \TOTOF to perform PID on the track
itself. 
This means that, in principle, each track  has to be removed by the
sample before calculating the \TOTOF, repeating this procedure for each track.
This approach would result in an excessive request of computing resources when the number of tracks is large.
Therefore, in order to optimize the procedure, the tracks are divided into ten momentum intervals. The \TOTOF is
calculated for each momentum interval using only the tracks belonging
to the other nine momentum bins. With this procedure the \TOTOF to be
used in Eq.~\ref{eq:pidres} to perform PID on a track is not biassed
by the implicit identification of the track performed by the \tev algorithm
with the TOF and is evaluated
using only the tracks in the momentum bins other than the one the track belongs to.
Finally, to avoid an excessive computational load due to the combinatorics, this
evaluation is done dividing the sample of tracks in the event in
several subsamples and the
weighted average of the results is then taken. 

It should be noted that \sigmaTOTOF is dependent on the event
track multiplicity because, according to Eq.~\ref{eq:sigTOF} it scales as
$\sim 1/\sqrt{n_{\rm{tracks}}}$. 

\subsection{Event collision time measurement performed by the T0 detector}
The T0 detector can provide two time measurements,
$t_{\rm{T0A}}$ and $t_{\rm{T0C}}$, one for each of its two sub-detectors T0A and
T0C, corresponding to the fastest signals among its photomultipliers. 
When both values are available, the event collision time is
defined as \TOTOAC=$(t_{\rm{T0A}}+t_{\rm{T0C}})/2$, which is independent of the event vertex position.
In low multiplicity events, when only one of the two arrays of
Cherenkov counters produces a signal, $t_{\rm{T0A}}$ or $t_{\rm{T0C}}$ can be
used as a measurement of the event collision time once a correction for the $z$-position of the primary vertex (as measured by the ITS with an accuracy of ~50~$\mu$m) is taken into account.

The time resolution of the T0 detector~\cite{aliceperf} is related to the number of photoelectrons emitted from the photocathode of each PMT. This, in turn, is directly proportional to the number of MIPs traversing the quartz radiator. In principle it would be possible to estimate the resolution for each event based on the registered amplitude in each T0 module but the analysis procedures implemented during \run{1} yielded only the average value per run.
As a consequence the time resolution depends on the average multiplicity of the events
in the run and hence on the colliding system. At the moment, the small
dependence of \sigmaTOTO on the track multiplicity is not taken into
account since it is only of the order of a maximum of 20\%,
negligible when
compared to the dependence of \sigmaTOTOF on the TOF track multiplicity as will be shown later, and
smaller than the run by run fluctuation. When both $t_{\rm{T0A}}$ and
$t_{\rm{T0C}}$ measurements are available the resolution can be estimated by the width of
the $(t_{\rm{T0A}}-t_{\rm{T0C}})/2$ distribution after both $t_{\rm{T0A}}$ and
$t_{\rm{T0C}}$  are corrected for the vertex position.  In \PbPb and \pp
collisions the resolutions are
\sigmaTOTOAC$\sim$25 and \sigmaTOTOAC$\sim$50 ps respectively. The
difference is due to the different average multiplicity of the events
in the two colliding systems
and the resulting different signal amplitudes. When only $t_{\rm{T0A}}$ or
$t_{\rm{T0C}}$ are available, the resolutions are \sigmaTOTOA$\sim$50 ps
and \sigmaTOTOC$\sim$30 ps in \PbPb collisions and \sigmaTOTOA$\sim$100 ps
and \sigmaTOTOC$\sim$60 ps in \pp collisions. The difference is
due to the different distance of T0A and T0C from the interaction
point.

To reach this time resolution, an accurate calibration procedure for
T0 is needed. Before every data taking period, gain and slewing corrections
are determined  using a set of laser runs, where the laser intensity
is varied.  The mean time value for each photomultiplier, after
slewing correction, is optimized for the minimum bias trigger for each
run. 

\subsection{Combination of the TOF and T0 measurements}
For each event, \tev is obtained
combining in a single estimation (\TOBest) the results from the different methods available.

If the \tev  measurement can be provided by only TOF or T0 detector,
\TOBest will correspond respectively to \TOTOF or \TOTO.  
If both of them are available than \TOBest is estimated by their
weighted mean where the weights are the inverse of the square of the resolutions. 
If both methods are not available, \TOBest fails and \tev is defined by the \TOFILL. In
the last case, the resolution is $\sim$200 ps.

The relative occurrence and resolutions of these three cases depend on the multiplicity of the event and therefore, indirectly, on the collision type, as will be shown in 
Sec.~\ref{sec:results}.

\begin{figure}[t!]
\centering
\includegraphics[width=10.0cm]{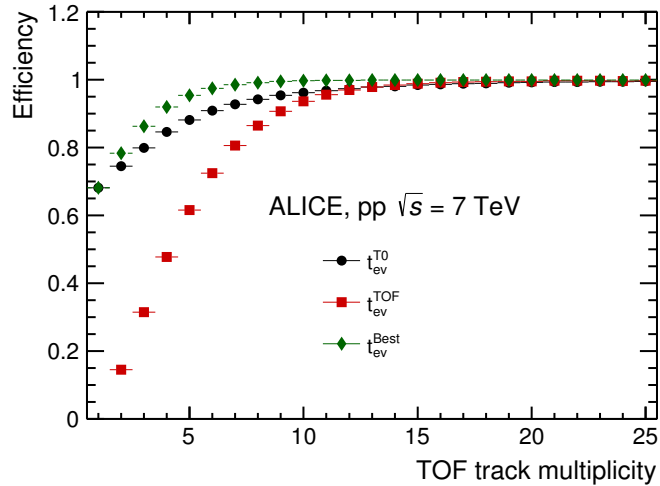}
\caption{Efficiency of the \TOTO (circles), \TOTOF (squares) and
\TOBest (diamond) as a function of the TOF track multiplicity for \pp
collisions at \sqrts~=~7~TeV.}
\label{fig:Eff_plotpp}
\end{figure}

\begin{figure}[t!]
\centering
\includegraphics[width=10.0cm]{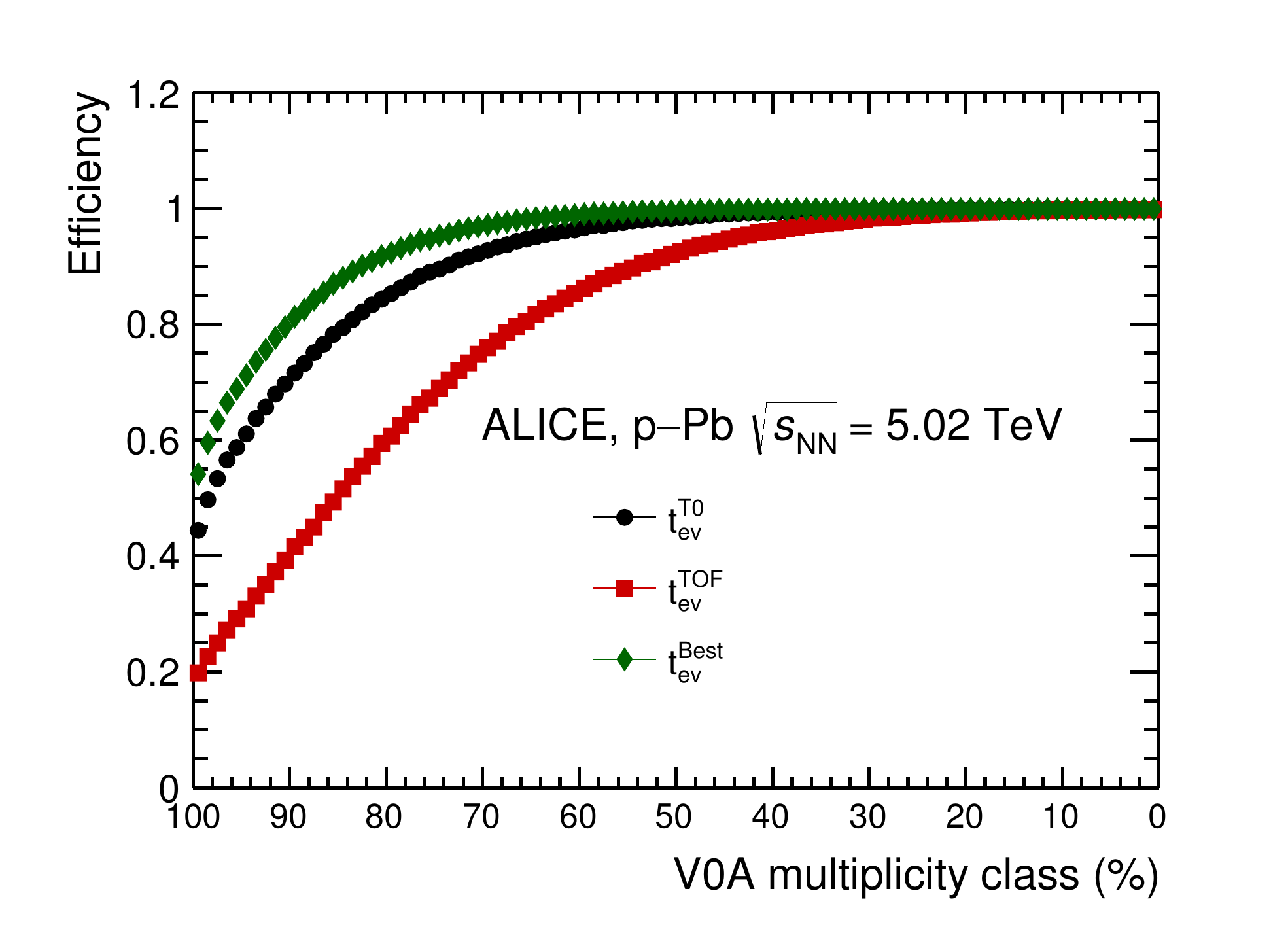}
\includegraphics[width=10.0cm]{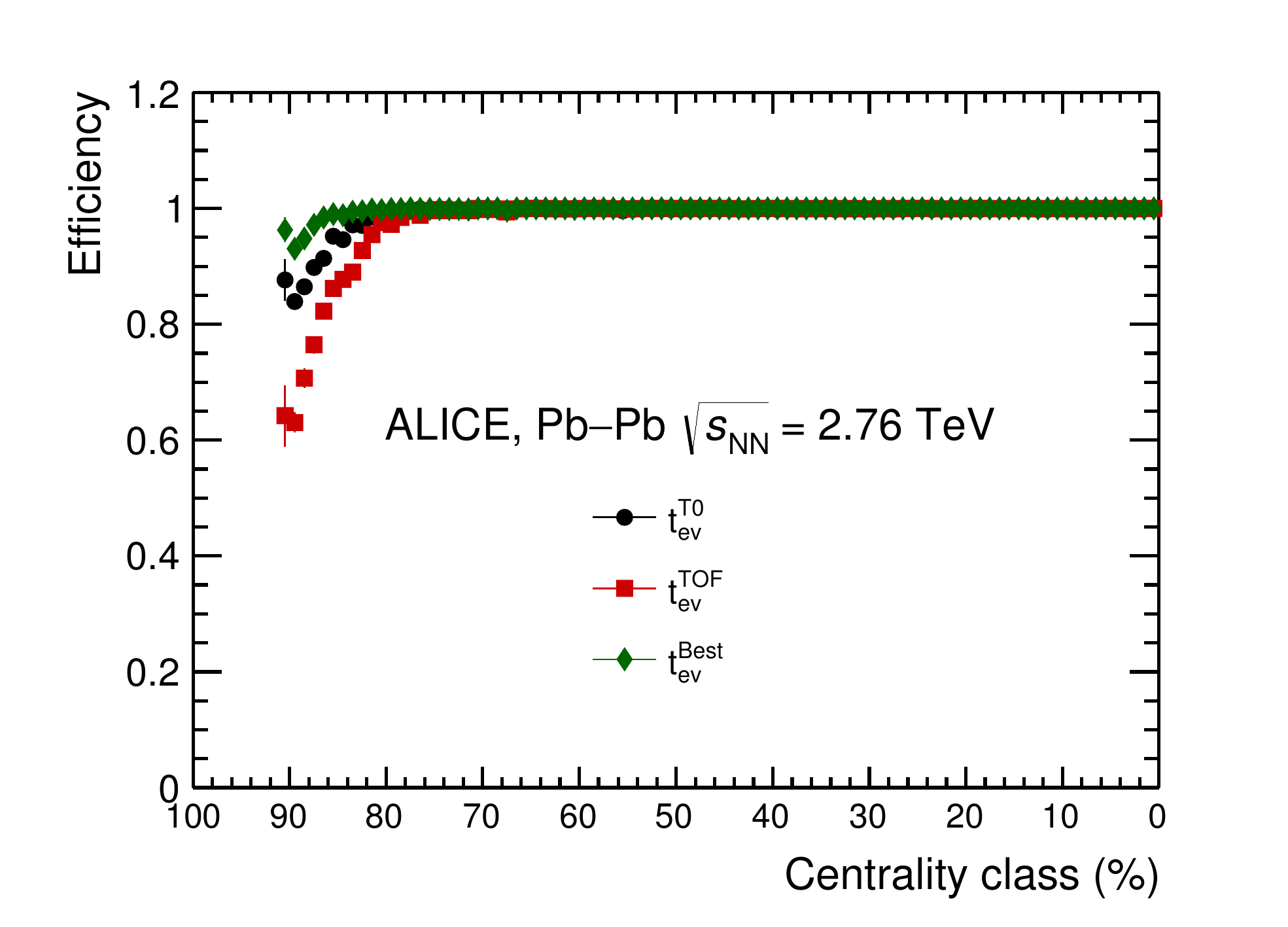}
\caption{Efficiencies of the methods \TOTO (circles), \TOTOF (squares) and
\TOBest (diamond) as a function of the V0A multiplicity class for p-Pb
collisions at \sqrts~=~2.76~TeV (top) and of the V0M Centrality class for \PbPb collisions\sqrtsNN~=~2.76~TeV (bottom).}
\label{fig:Eff_plotpPbCentr}
\end{figure}

\section{Results}
\label{sec:results}
Results related to the efficiency of the methods used to define
the event collision time as a function of the TOF track multiplicity,
their resolution and their impact on the PID performance are reported in this section. For \pPb and
\PbPb collision systems the analysis is provided also as a function
of the multiplicity or centrality of the collision.

\subsection{Efficiency of  the determination of \TOTOF, \TOTO and \TOBest}
In Figure~\ref{fig:Eff_plotpp} the efficiency of the determination of \TOTOF, \TOTO and
\TOBest is reported as a function of the TOF track multiplicity
in \pp collisions at $\sqrt{s}$~=~7~TeV. 

The efficiency is defined as the fraction of events for which the
\TOTOF, \TOTO or \TOBest has been measured compared to the ones selected as
explained in Sec.~\ref{sec:intro}. 
Since \TOTOF and thus \TOBest are defined in ten momentum bins
(see Sec.~\ref{sec:TOFmethods}) they are considered efficient if 
the measurement is available in at least one momentum bin. 

The TOF track multiplicity of the event is the number of tracks
matched with a hit on the TOF detector that is the number of tracks
with an associated \tof measurement. This is the minimal request for a
track to be identified by the \tof method. It is important to notice
that the TOF track multiplicity
does not represent the number of tracks that are used by the 
TOF algorithm to compute the \TOTOF, that is actually slightly lower
since in the algorithm a further basic
selection on the quality of the track is applied to guarantee a good
quality of the \TOTOF. What is reported in Fig.~\ref{fig:Eff_plotpp}
is, therefore, not the algorithmic efficiency. 

From Sec.~\ref{sec:TOFmethods} it is evident that
the minimum number of tracks to compute \TOTOF is two. Therefore the
\TOTOF efficiency in the first bin is not shown in Fig.~\ref{fig:Eff_plotpp}. 

In \pp collisions, for very low multiplicity events,
the T0 detector can provide a \tev measurement
with an efficiency of the order of $\sim$70\% that increases with the
track multiplicity.
At the same time, for all events having high multiplicity, the \TOTOF method
is able to provide a \tev measurement.

The curve corresponding to \TOBest shows how the two techniques can be combined
to minimize the number of events, in particular at low multiplicity,
where an event-by-event \tev measurement cannot be provided and only
\TOFILL is available.
In \pp collisions at \sqrts~=~7~TeV, when more than three tracks reach
the TOF the event time efficiency is greater than 80\%.

In Figure~\ref{fig:Eff_plotpPbCentr} the efficiency of the \TOTO, \TOTOF and
\TOBest is reported as a function of the V0A multiplicity class  in
\pPb and centrality in \PbPb collisions, respectively.

\begin{table}[t!]
\renewcommand\arraystretch{1.5} 
\centering
\begin{tabular}{|c|c|c|c|c|}
\hline
 & $\mathbf{t_{ev}^{TOF}} (\%)$ &
                                  \multicolumn{3}{c|}{$\mathbf{t_{ev}^{T0}}$ (\%)} \\
\hline
\hline
& & $\mathbf{t_{ev}^{T0A}}$ & $\mathbf{t_{ev}^{T0C}}$ & $\mathbf{t_{ev}^{T0AC}}$ \\
\hline
\hline
pp \sqrts~=~7 TeV& 52.5 & 18.0& 21.8& 45.2 \\
\hline 
&52.5 & \multicolumn{3}{c|}{$\sum$=85.0} \\
\hline
\pPb \sqrtsNN~=~5.02 TeV& 81.8 & 13.0 & 11.0 & 68.4 \\
\hline
&81.8 & \multicolumn{3}{c|}{$\sum$=92.4} \\
\hline
\PbPb \sqrtsNN~=~2.76 TeV& 99.6 & 0.3 & 0.5 & 98.9\\
\hline
&99.6 & \multicolumn{3}{c|}{$\sum$=99.7} \\
\hline
%\hline
\end{tabular}
\caption{Fraction of events (percentage) for which the \TOTOF and \TOTO  can be provided when explicitly requested. The results are
  shown for \pp collisions at \sqrts~=~7~TeV, \pPb collisions at
  \sqrtsNN~=~5.02~TeV and \PbPb collisions at \sqrtsNN~=~2.76~TeV.}
\label{tab:eff_TOFMatch1}
\end{table}

\begin{table}[th!]
\renewcommand\arraystretch{1.5} 
%\begin{sidewaystable}
\centering
\begin{tabular}{|c|c|c|c|c|c|c|c|c|}
\hline
  &  \multicolumn{7}{c|}{$\mathbf{t_{ev}^{Best}}$
                                                                                      (\%)}
  &\\
\hline
\hline
&
                    $\mathbf{t_{ev}^{TOF}}$ & $\mathbf{t_{ev}^{T0A}}$ & $\mathbf{t_{ev}^{TOF+T0A}}$
                   & $\mathbf{t_{ev}^{T0C}}$ & $\mathbf{t_{ev}^{TOF+T0C}}$
                                                     & $\mathbf{t_{ev}^{T0AC}}$
                   & $\mathbf{t_{ev}^{TOF+T0AC}}$ &  $\mathbf{t_{ev}^{Fill}}$ \\
\hline
\hline
pp \sqrts~=~7 TeV& 4.0 & 10.8 & 7.2 & 11.5 & 10.3
                   & 14.2 & 31.0 & 11.0  \\
\hline 
&  \multicolumn{7}{c|}{$\sum$=89.0} & \\
\hline
\pPb \sqrtsNN~=~5.02 TeV& 2.9 & 4.2 & 8.8 & 4.0 & 7.0 &
                                                                      5.4
                   & 63.0 & 4.7 \\
\hline
&  \multicolumn{7}{c|}{$\sum$=95.3} & \\
\hline
\PbPb \sqrtsNN~=~2.76 TeV& 0.2 & 0.09 & 0.2 & 0.1 & 0.4 &
                                                                    0.1
                   & 98.8 & 0.1 \\
\hline
& \multicolumn{7}{c|}{$\sum$=99.9} & \\
\hline
\end{tabular}
\caption{Fraction of events (percentage) for which the
  \TOBest can be provided when explicitly requested (total and for
  each subcase). The results are
  shown for \pp at \sqrts~=~7~TeV, \pPb collisions at
  \sqrtsNN~=~5.02~TeV and \PbPb collisions at \sqrtsNN~=~2.76~TeV.}
\label{tab:eff_TOFMatch2}
\end{table}

In \pPb collisions, from 0 to 40\% V0A multiplicity class, both the T0
and the
TOF are fully efficient in determining the collision time. For more peripheral events
the T0 detector has the highest efficiency in providing a \tev measurement.
For \PbPb collisions only for the most peripheral events (centrality
$>$ 80\%) the T0 has an efficiency higher than the TOF. 
In \PbPb collisions the \TOBest is 100\%
efficient except for the very peripheral events. As a consequence,
the \TOFILL is basically never used. It is worth to notice that the
efficiency curves would have similar trend than the ones in
Fig.~\ref{fig:Eff_plotpp} once plotted as a function of the TOF track
multiplicity instead of the V0A multiplicity class or centrality since the efficiency mainly
depends on the track multiplicity. 

The overall efficiency defined as the
fraction (in percentage) of events with at least one
track associated to a hit in the TOF detector for which the \TOTO,
\TOTOF and \TOBest can be provided, is
reported in Table~\ref{tab:eff_TOFMatch1} and Table~\ref{tab:eff_TOFMatch2}.
The first column of Table~\ref{tab:eff_TOFMatch1} represents the fraction of events (in \%) for which the \TOTOF
can be provided in at least one momentum bin. It can be
seen that in \pp at \sqrts~=~7~TeV \TOTOF is measured only in less
than 53\% of events. This percentage increases reaching 99.6\% in \PbPb collisions.
The second column shows the fraction of events (in \%) for which the \TOTO can
be provided. In this case, if both T0A and T0C provide a signal, 
the \TOTOAC is used otherwise the individual \TOTOA or \TOTOC are used. From \pp at
\sqrts~=~7~\TeV to \PbPb  at \sqrtsNN~=~2.76~\TeV the
efficiency of the \TOTO increases from 85\% to 99.7\% with an increase
of the efficiency of the \TOTOAC as expected. The fraction of events
for which only the T0A or T0C is used decreases.

In Table~\ref{tab:eff_TOFMatch2} the efficiency (in \%) of the \TOBest
also for each exclusive subcases
is reported. 
The outcomes of the possible combinations resulting in a \TOBest
measurement are detailed in the seven subcolumns. 
In \PbPb collisions,
for most of the events both \TOTOF and \TOTO are available.

\begin{figure}[th!]
\centering
\includegraphics[width=10.0cm]{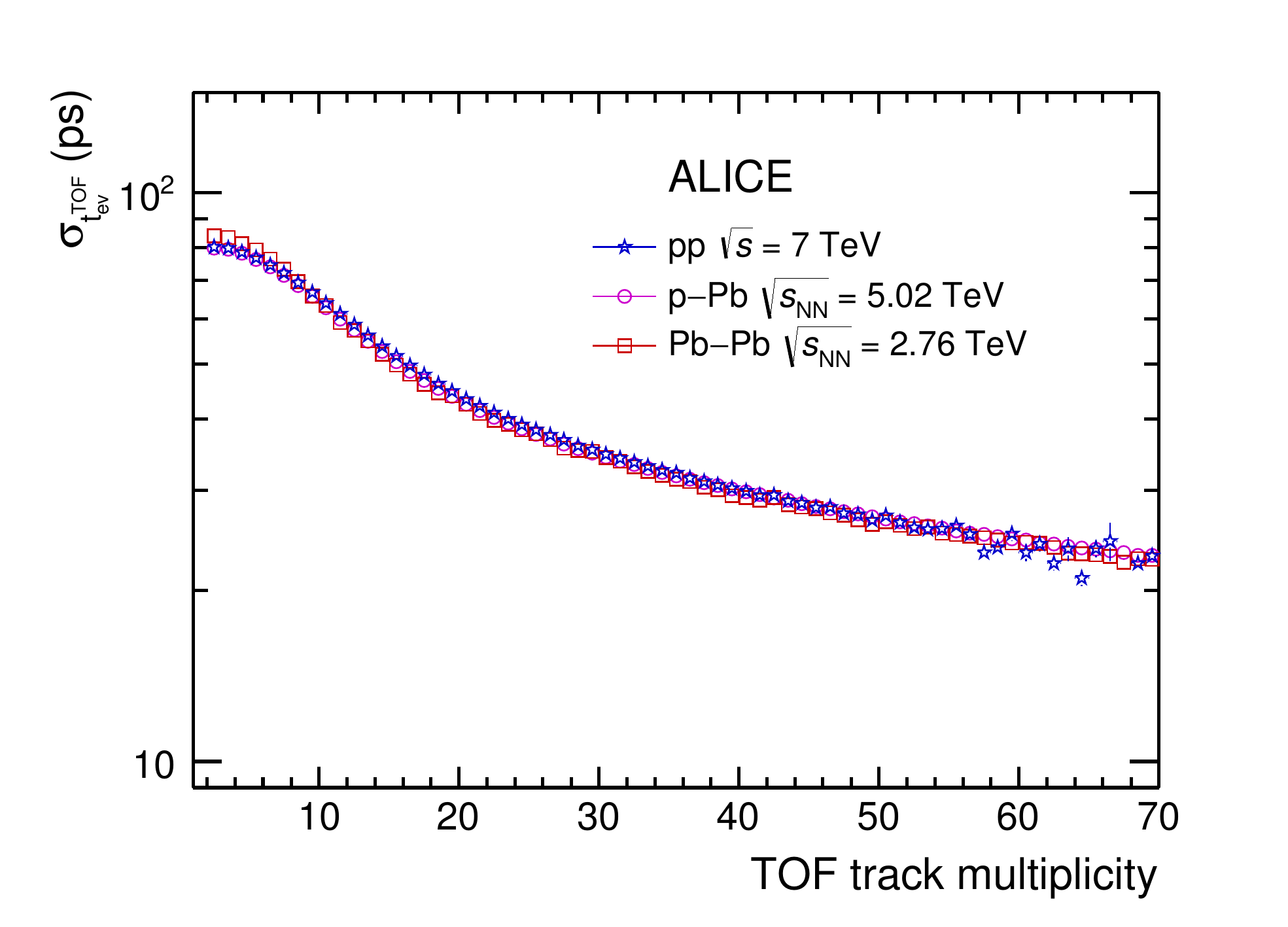}
\includegraphics[width=10.0cm]{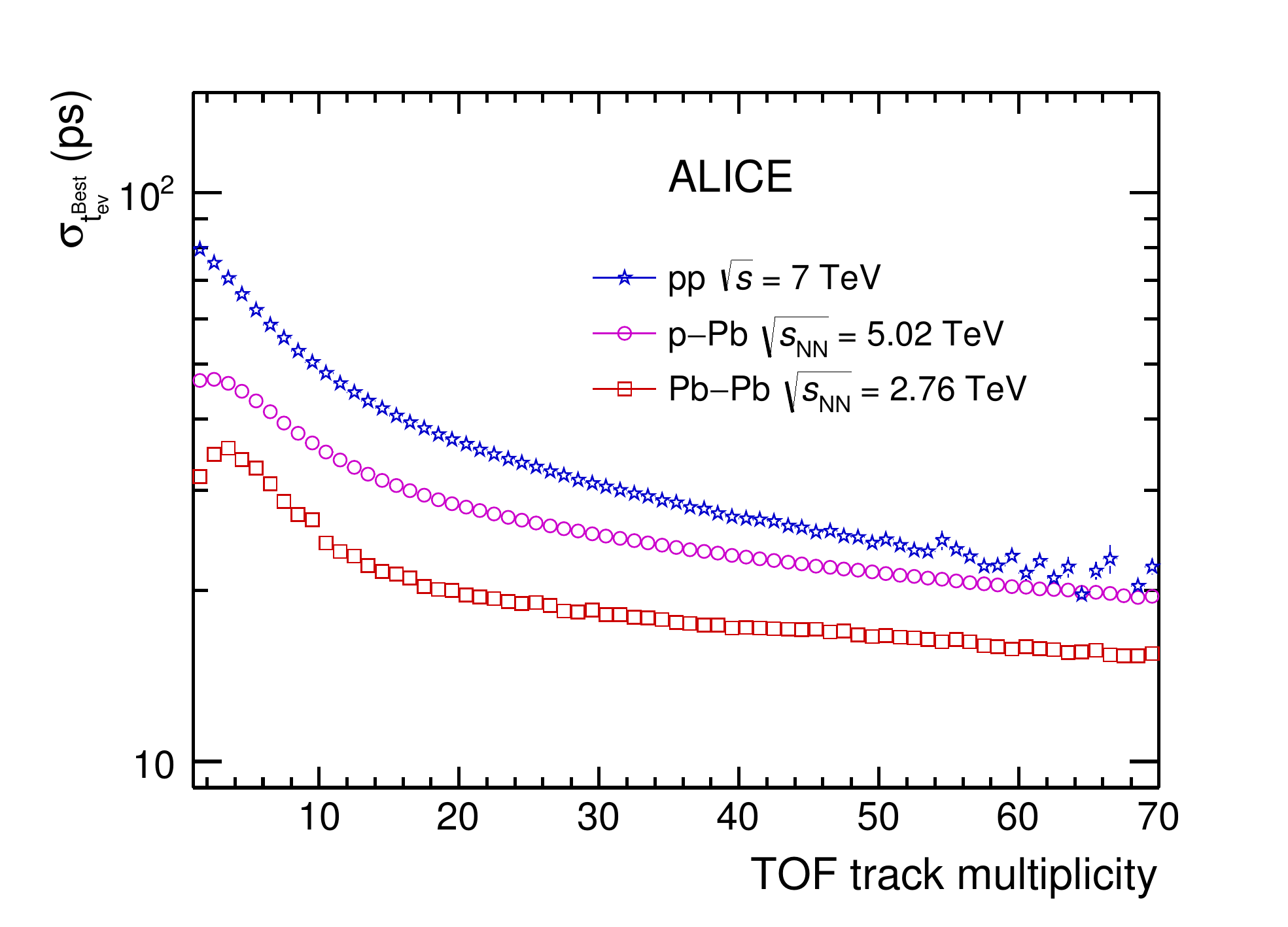}
\caption{Resolution of \TOTOF (top) and \TOBest (bottom) as a function of the TOF track multiplicity for \pp
  collisions at \sqrts~=~7~TeV (star), \pPb collisions at \sqrtsNN~=~5.02~TeV (circle) and \PbPb collisions
at \sqrtsNN~=~2.76~TeV (square). As a reference, TOF track
multiplcity=15 corresponds to 50\% V0A multiplicity class in \pPb and
80\% centrality class in \PbPb.}
\label{fig:T0TOF_Reso}
\end{figure}

\subsection{Resolution of the \TOTOF and \TOBest as a function of the TOF track multiplicity}
In Figure~\ref{fig:T0TOF_Reso} (top) the \TOTOF resolution (\sigmaTOTOF) is shown
as a function of the TOF track multiplicity for \pp data at \sqrts~=~7~\TeV, \pPb collisions at \sqrtsNN~=~5.02~\TeV and \PbPb collisions
at \sqrtsNN~=~2.76~\TeV. 

The trend with the multiplicity is the same for all
the data sets since \sigmaTOTOF mainly depends on the number of
tracks used by the algorithm that is related in turn to the
TOF track multiplicity.

The resolution improves from $\sim$80 ps in low multiplicity events,
to 20 ps for high multiplicity events. 
As a consequence, \sigmaTOTOF is a significant contribution of the TOF
PID resolution $\sigma_{PID}$ reported in Eq.~\ref{eq:sigmapid} only
for low multiplicity events, when it is of the same order of the
TOF resolution $\sigma_{t_{TOF}}$. It becomes negligible at higher track
multiplicities. While the resolution as a function of multiplicity is the same for the different
colliding systems, it is important to remind here that what is different
is the overall fraction of events for which the
\TOTOF can be provided as can be seen in Table~\ref{tab:eff_TOFMatch1}.
It depends on the mean multiplicity of the
events that increases from \pp to \pPb and to \PbPb collisions.

In Figure~\ref{fig:T0TOF_Reso} (bottom) the resolution of \TOBest
(\sigmaTOBest) is reported as a function of the TOF track multiplicity for \pp
collisions at \sqrts~=~7~\TeV,  for  \pPb and \PbPb data at
\sqrtsNN~=~5.02~\TeV and \sqrtsNN~=~2.76~\TeV, respectively.
It depends on two main factors: the track
multiplicity and the colliding system. The first defines the \TOTOF
resolution while the second the \sigmaTOTO that decreases moving
from \pp to \pPb to \PbPb since, as explained before,
\sigmaTOTO depends only on the mean event multiplicity being defined
per run and not per event. 
The exclusive probability of the seven possible subcases of \TOBest plays a role here in particular to explain the pattern observed at low multiplicity in Fig.~\ref{fig:T0TOF_Reso} for the \PbPb case. 

\begin{figure}[t!]
\centering
\includegraphics[width=10.0cm]{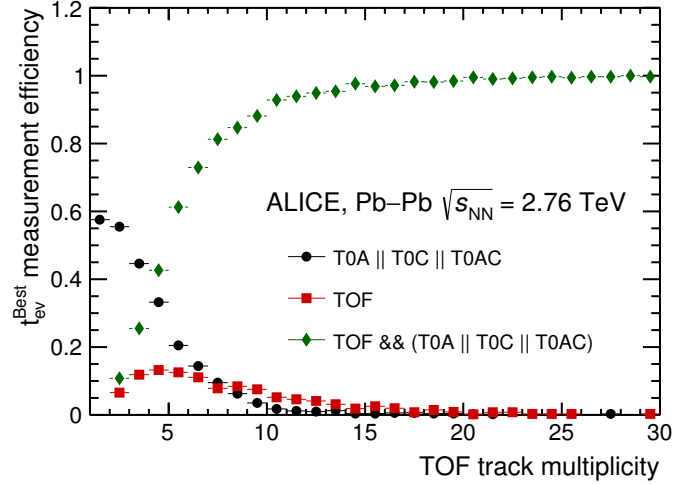}
\caption{Fraction of events for which the \TOBest is provided
  exclusively by
  the T0, no matter if T0A, T0C or T0AC (circle), or exclusively by the TOF (square)
  or a combination of the two (green) in \PbPb collisions at
  \sqrtsNN~=~2.76~TeV.}
\label{fig:Eff_plotBestPbPb}
\end{figure}

In Figure~\ref{fig:Eff_plotBestPbPb} the efficiency as a function of the TOF
track multiplicity of the possible outcomes of the \TOBest are shown
for \PbPb collisions at \sqrtsNN~=~2.76~TeV. 

It is evident
that, for less than 3 tracks matched to the TOF, for most of the events
the \TOBest is provided by the T0 while, increasing the multiplicity,
the combination of the T0 and TOF measurements becomes the dominant
term.
The interplay of all these factors define the shape of the
\sigmaTOBest reported in the bottom plot of Fig.~\ref{fig:T0TOF_Reso}.

\subsection{Effect of the \tev resolution on the PID performance}
In this section, the impact on  the PID performance due to the
different methods used for the event collision time determination is
assessed. This is studied via the K-$\pi$ and p-K
separation power: $n\sigma_{i,j} (t_{\rm{ev}}^{k}) = (t_{\rm{exp},i} -
t_{\rm{exp},j}) / \sigma_{\rm{PID},j}(t_{\rm{ev}}^{k})$ where
$i,j=\pi,$ K, p and 
$\sigma^{2}_{\rm{PID},j}(t_{\rm{ev}}^{k})=\sigma^{2}_{\rm{TOF}}+\sigma^{2}_{t_{\rm{ev}}}+\sigma^{2}
_{t_{\rm{exp},j}}$ with $k$= TOF, T0, Best and Fill.

\begin{figure}[t!]
\centering
\includegraphics[width=10.0cm]{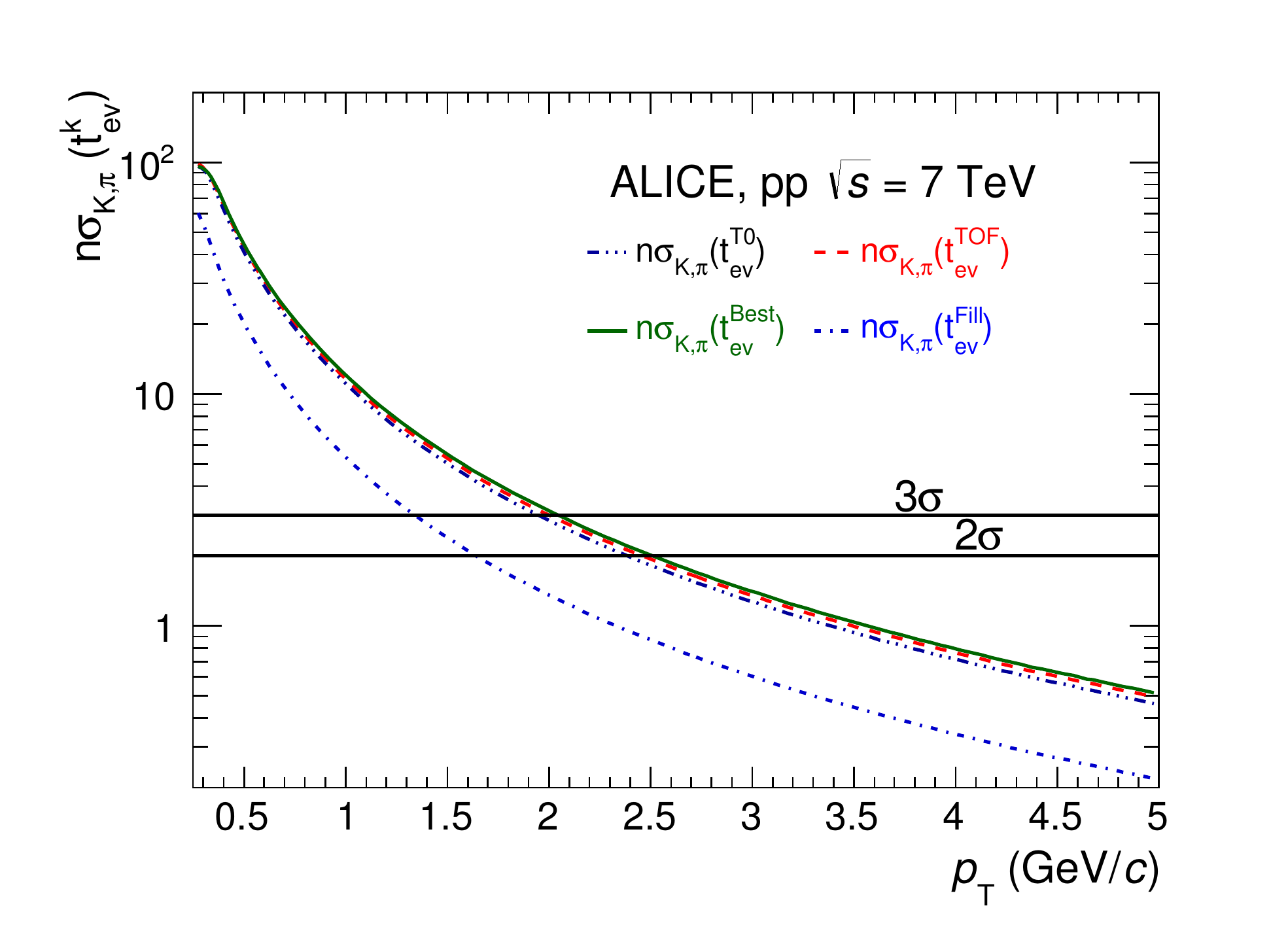}
\includegraphics[width=10.0cm]{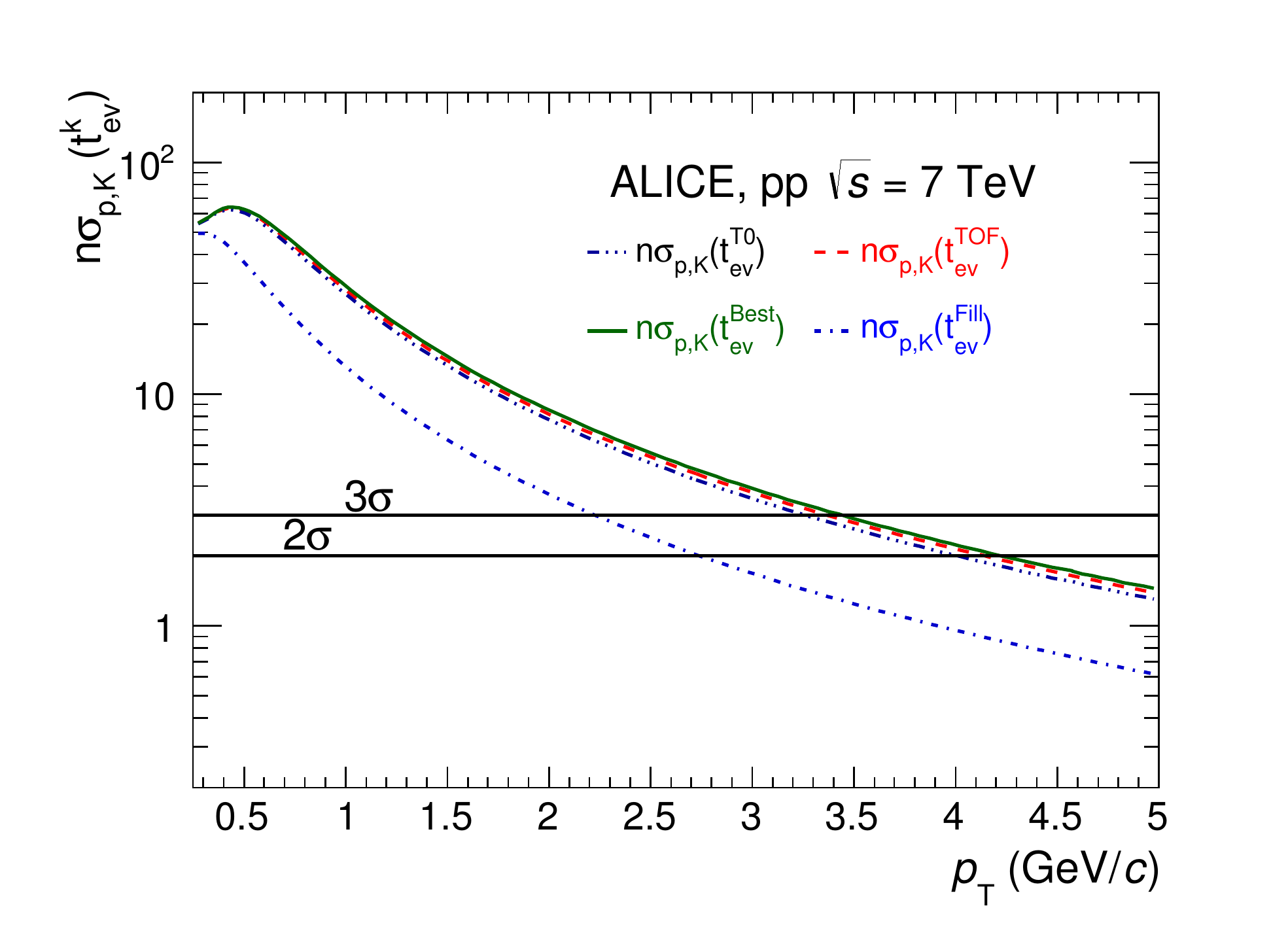}
\caption{K-$\pi$ (top) and p-K (bottom) separation power as a function
of the transverse momentum of a track when \TOTOF (dashed line), \TOTO
(dash-dottet line),
\TOBest (solid line) and \TOFILL (dotted line) are
used.}
\label{fig:Sigma_Trk7TeV}
\end{figure}

In Figure~\ref{fig:Sigma_Trk7TeV}, $n\sigma_{\rm{K,\pi}}(t_{\rm{ev}}^{k})$ and $n\sigma_{\rm{p,K}}(t_{\rm{ev}}^{k})$ are
shown as a function of the transverse momentum of the track.

The separation power does not significantly change when changing the
\tev estimator (\TOTOF, \TOTO or \TOBest). On the other hand, it gets worse if the \TOFILL is used 
since its resolution is much worse than the one of all the others. If a three sigma separation is requested, the $\pi$-K separation is achievable
only up to 1.3 \gmom ~ instead of up to 2 \gmom~if the \TOFILL is
used and the K-p separation can be defined
only up to 2.2 \gmom ~ instead of up to 3.5 \gmom. 

\section{Conclusions}
\label{concl}

The determination of the event collision time in ALICE is needed to perform particle identification in the
intermediate region of momentum (0.5-4.0 \gevc) with the \tof method.
It can be provided on an event-by-event basis by the T0 detector
(\TOTO)  or the TOF detector itself
(\TOTOF). When both the measurements are available a weighted
mean can be defined (\TOBest).
In case none of the previous methods can be used, mainly for
low multiplicity events, only an average collision time (\TOFILL) can be
considered, with a resolution of $\sim$200 ps, which worsens the TOF PID performance.  
In this paper the methods for the event collision time determination
in ALICE have
been reviewed, together with their performance during LHC
\run{1} data in terms of efficiency, resolution and impact on the TOF PID.

It has been shown how, for very low multiplicity events,
the T0 detector plays a crucial role since it has a higher
efficiency in providing \tev when compared to the TOF detector. For
example, when 
five tracks reach the TOF, the \TOTO efficiency is $\sim$85\% compared to
the 60\% of the TOF detector. The \TOTOF efficiency increases with the rise of the
track multiplicity reaching $\sim$100\% when 15 tracks reach the TOF.

In the analysed data set and given the current level of calibration of detectors, for high multiplicity events the resolution of the event collision time becomes a negligible term in the time-of-flight resolution. This is achieved combining the \TOTOF and \TOTO measurements.
In \pp collisions at \sqrts~=~7~TeV only for
the 52.5\% of events with at least one track associated to a hit on
the TOF detector the \TOTOF
can be provided. In \pPb collisions this fraction increases to 81.8\%
reaching 99.6\% in \PbPb collisions.

To increase the PID performance it is important to use the
\TOBest which combines the high \TOTO efficiency at low multiplicity
events with the better \TOTOF resolution at high multiplicity events.
Finally, the impact of the method used for
the event collision time determination on the TOF PID performance has
been discussed, showing how it gets better when \tev is computed event-by-event  improving for example a three sigma $\pi$-K separation from 1.3 \gevc to 2 \gevc with respect to when the \TOFILL has to be used.

%%%%% acknowledgements
\newenvironment{acknowledgement}{\relax}{\relax}
\begin{acknowledgement}
\section*{Acknowledgements}
% Version: 2016-09-13

The ALICE Collaboration would like to thank all its engineers and technicians for their invaluable contributions to the construction of the experiment and the CERN accelerator teams for the outstanding performance of the LHC complex.
The ALICE Collaboration gratefully acknowledges the resources and support provided by all Grid centres and the Worldwide LHC Computing Grid (WLCG) collaboration.
The ALICE Collaboration acknowledges the following funding agencies for their support in building and running the ALICE detector:
A. I. Alikhanyan National Science Laboratory (Yerevan Physics Institute) Foundation (ANSL), State Committee of Science and World Federation of Scientists (WFS), Armenia;
Austrian Academy of Sciences and Nationalstiftung f\"{u}r Forschung, Technologie und Entwicklung, Austria;
Conselho Nacional de Desenvolvimento Cient\'{\i}fico e Tecnol\'{o}gico (CNPq), Financiadora de Estudos e Projetos (Finep), Funda\c{c}\~{a}o de Amparo \`{a} Pesquisa do Estado de S\~{a}o Paulo (FAPESP) and , Brazil;
Ministry of Education of China (MOEC)), Ministry of Science \& Technology of China (MSTC) and National Natural Science Foundation of China (NSFC), China;
Ministry of Science, Education and Sport and Croatian Science Foundation, Croatia;
Ministry of Education, Youth and Sports of the Czech Republic, Czech Republic;
The Danish Council for Independent Research | Natural Sciences, Danish National Research Foundation (DNRF) and The Carlsberg Foundation, Denmark;
Helsinki Institute of Physics (HIP), Finland;
Institut National de Physique Nucl\'{e}aire et de Physique des Particules (IN2P3) and Centre National de la Recherche Scientifique (CNRS) and Commissariat \`{a} l'Energie Atomique (CEA), France;
Bundesministerium f\"{u}r Bildung, Wissenschaft, Forschung und Technologie (BMBF) and GSI Helmholtzzentrum f\"{u}r Schwerionenforschung GmbH, Germany;
Ministry of Education, Research and Religious Affairs, Greece;
National Research, Development and Innovation Office, Hungary;
Department of Atomic Energy Government of India (DAE), India;
Indonesian Institute of Science, Indonesia;
Istituto Nazionale di Fisica Nucleare (INFN) and Centro Fermi - Museo Storico della Fisica e Centro Studi e Ricerche Enrico Fermi, Italy;
Japanese Ministry of Education, Culture, Sports, Science and Technology (MEXT), Japan Society for the Promotion of Science (JSPS) KAKENHI and Institute for Innovative Science and Technology , Nagasaki Institute of Applied Science (IIST), Japan;
Consejo Nacional de Ciencia (CONACYT) y Tecnolog\'{i}a, through Fondo de Cooperaci\'{o}n Internacional en Ciencia y Tecnolog\'{i}a (FONCICYT) and Direcci\'{o}n General de Asuntos del Personal Academico (DGAPA), Mexico;
Nationaal instituut voor subatomaire fysica (Nikhef), Netherlands;
The Research Council of Norway, Norway;
Commission on Science and Technology for Sustainable Development in the South (COMSATS), Pakistan;
Pontificia Universidad Cat\'{o}lica del Per\'{u}, Peru;
Ministry of Science and Higher Education and National Science Centre, Poland;
Korea Institute of Science and Technology Information and National Research Foundation of Korea (NRF), Republic of Korea;
Romanian National Agency for Science, Technology and Innovation and Ministry of Education and Scientific Research, Institute of Atomic Physics, Romania;
Ministry of Education and Science of the Russian Federation, National Research Centre Kurchatov Institute and Joint Institute for Nuclear Research (JINR), Russia;
Ministry of Education, Science, Research and Sport of the Slovak Republic, Slovakia;
National Research Foundation of South Africa, South Africa;
Centro de Investigaciones Energ\'{e}ticas, Medioambientales y Tecnol\'{o}gicas (CIEMAT), Centro de Aplicaciones Tecnol\'{o}gicas y Desarrollo Nuclear (CEADEN), Cubaenerg\'{\i}a, Cuba and Ministerio de Ciencia e Innovacion, Spain;
Swedish Research Council (VR) and Knut \& Alice Wallenberg Foundation (KAW), Sweden;
European Organization for Nuclear Research, Switzerland;
Suranaree University of Technology (SUT), National Science and Technology Development Agency (NSDTA) and Office of the Higher Education Commission under NRU project of Thailand, Thailand;
Turkish Atomic Energy Agency (TAEK), Turkey;
National Academy of  Sciences of Ukraine, Ukraine;
Science and Technology Facilities Council (STFC), United Kingdom;
United States Department of Energy, Office of Nuclear Physics (DOE NP) and National Science Foundation of the United States of America (NSF), United States of America.    %%%%%%% done by webmaster team
\end{acknowledgement}

\bibliographystyle{utphys}
\bibliography{paper}

%%%%%%%%% appendix with author list
\newpage
\appendix
\section{The ALICE Collaboration}
\label{app:collab}

% Collaboration: CERN-LHC-ALICE
% Generation Date is 2016-09-13

% How to use:
%%%%%%%%% appendix with author list
%\appendix
%\section{The ALICE Collaboration}
%\label{app:collab}
%\input{authors-list.tex}  %%%%%%% get the latest version before submitting

\begingroup
\small
\begin{flushleft}
J.~Adam$^\textrm{\scriptsize 39}$,
D.~Adamov\'{a}$^\textrm{\scriptsize 86}$,
M.M.~Aggarwal$^\textrm{\scriptsize 90}$,
G.~Aglieri Rinella$^\textrm{\scriptsize 35}$,
M.~Agnello$^\textrm{\scriptsize 113}$\textsuperscript{,}$^\textrm{\scriptsize 31}$,
N.~Agrawal$^\textrm{\scriptsize 48}$,
Z.~Ahammed$^\textrm{\scriptsize 137}$,
S.~Ahmad$^\textrm{\scriptsize 18}$,
S.U.~Ahn$^\textrm{\scriptsize 70}$,
S.~Aiola$^\textrm{\scriptsize 141}$
% ...
\renewcommand\labelenumi{\textsuperscript{\theenumi}~}

\section*{Affiliation notes}
\renewcommand\theenumi{\roman{enumi}}
\begin{Authlist}
\item \Adef{0}Deceased
\item \Adef{idp1803344}{Also at: Georgia State University, Atlanta, Georgia, United States}
\item \Adef{idp3218624}{Also at: Also at Department of Applied Physics, Aligarh Muslim University, Aligarh, India}
\item \Adef{idp3967088}{Also at: M.V. Lomonosov Moscow State University, D.V. Skobeltsyn Institute of Nuclear, Physics, Moscow, Russia}
\end{Authlist}

\section*{Collaboration Institutes}
\renewcommand\theenumi{\arabic{enumi}~}

$^{1}$A.I. Alikhanyan National Science Laboratory (Yerevan Physics Institute) Foundation, Yerevan, Armenia
\\
$^{2}$Benem\'{e}rita Universidad Aut\'{o}noma de Puebla, Puebla, Mexico
\\
$^{3}$Bogolyubov Institute for Theoretical Physics, Kiev, Ukraine
\\
$^{4}$Bose Institute, Department of Physics 
and Centre for Astroparticle Physics and Space Science (CAPSS), Kolkata, India
\\
$^{5}$Budker Institute for Nuclear Physics, Novosibirsk, Russia
\\
$^{6}$California Polytechnic State University, San Luis Obispo, California, United States
\\
$^{7}$Central China Normal University, Wuhan, China
\\
$^{8}$Centre de Calcul de l'IN2P3, Villeurbanne, Lyon, France
\\
$^{9}$Centro de Aplicaciones Tecnol\'{o}gicas y Desarrollo Nuclear (CEADEN), Havana, Cuba
\\
$^{10}$Centro de Investigaciones Energ\'{e}ticas Medioambientales y Tecnol\'{o}gicas (CIEMAT), Madrid, Spain
\\
$^{11}$Centro de Investigaci\'{o}n y de Estudios Avanzados (CINVESTAV), Mexico City and M\'{e}rida, Mexico
\\
$^{12}$Centro Fermi - Museo Storico della Fisica e Centro Studi e Ricerche ``Enrico Fermi', Rome, Italy
\\
$^{13}$Chicago State University, Chicago, Illinois, United States
\\
$^{14}$China Institute of Atomic Energy, Beijing, China
\\
$^{15}$Commissariat \`{a} l'Energie Atomique, IRFU, Saclay, France
\\
$^{16}$COMSATS Institute of Information Technology (CIIT), Islamabad, Pakistan
\\
$^{17}$Departamento de F\'{\i}sica de Part\'{\i}culas and IGFAE, Universidad de Santiago de Compostela, Santiago de Compostela, Spain
\\
$^{18}$Department of Physics, Aligarh Muslim University, Aligarh, India
\\
$^{19}$Department of Physics, Ohio State University, Columbus, Ohio, United States
\\
$^{20}$Department of Physics, Sejong University, Seoul, South Korea
\\
$^{21}$Department of Physics, University of Oslo, Oslo, Norway
\\
$^{22}$Department of Physics and Technology, University of Bergen, Bergen, Norway
\\
$^{23}$Dipartimento di Fisica dell'Universit\`{a} 'La Sapienza'
and Sezione INFN, Rome, Italy
\\
$^{24}$Dipartimento di Fisica dell'Universit\`{a}
and Sezione INFN, Cagliari, Italy
\\
$^{25}$Dipartimento di Fisica dell'Universit\`{a}
and Sezione INFN, Trieste, Italy
\\
$^{26}$Dipartimento di Fisica dell'Universit\`{a}
and Sezione INFN, Turin, Italy
\\
$^{27}$Dipartimento di Fisica e Astronomia dell'Universit\`{a}
and Sezione INFN, Bologna, Italy
\\
$^{28}$Dipartimento di Fisica e Astronomia dell'Universit\`{a}
and Sezione INFN, Catania, Italy
\\
$^{29}$Dipartimento di Fisica e Astronomia dell'Universit\`{a}
and Sezione INFN, Padova, Italy
\\
$^{30}$Dipartimento di Fisica `E.R.~Caianiello' dell'Universit\`{a}
and Gruppo Collegato INFN, Salerno, Italy
\\
$^{31}$Dipartimento DISAT del Politecnico and Sezione INFN, Turin, Italy
\\
$^{32}$Dipartimento di Scienze e Innovazione Tecnologica dell'Universit\`{a} del Piemonte Orientale and INFN Sezione di Torino, Alessandria, Italy
\\
$^{33}$Dipartimento Interateneo di Fisica `M.~Merlin'
and Sezione INFN, Bari, Italy
\\
$^{34}$Division of Experimental High Energy Physics, University of Lund, Lund, Sweden
\\
$^{35}$European Organization for Nuclear Research (CERN), Geneva, Switzerland
\\
$^{36}$Excellence Cluster Universe, Technische Universit\"{a}t M\"{u}nchen, Munich, Germany
\\
$^{37}$Faculty of Engineering, Bergen University College, Bergen, Norway
\\
$^{38}$Faculty of Mathematics, Physics and Informatics, Comenius University, Bratislava, Slovakia
\\
$^{39}$Faculty of Nuclear Sciences and Physical Engineering, Czech Technical University in Prague, Prague, Czech Republic
\\
$^{40}$Faculty of Science, P.J.~\v{S}af\'{a}rik University, Ko\v{s}ice, Slovakia
\\
$^{41}$Faculty of Technology, Buskerud and Vestfold University College, Tonsberg, Norway
\\
$^{42}$Frankfurt Institute for Advanced Studies, Johann Wolfgang Goethe-Universit\"{a}t Frankfurt, Frankfurt, Germany
\\
$^{43}$Gangneung-Wonju National University, Gangneung, South Korea
\\
$^{44}$Gauhati University, Department of Physics, Guwahati, India
\\
$^{45}$Helmholtz-Institut f\"{u}r Strahlen- und Kernphysik, Rheinische Friedrich-Wilhelms-Universit\"{a}t Bonn, Bonn, Germany
\\
$^{46}$Helsinki Institute of Physics (HIP), Helsinki, Finland
\\
$^{47}$Hiroshima University, Hiroshima, Japan
\\
$^{48}$Indian Institute of Technology Bombay (IIT), Mumbai, India
\\
$^{49}$Indian Institute of Technology Indore, Indore, India
\\
$^{50}$Indonesian Institute of Sciences, Jakarta, Indonesia
\\
$^{51}$Inha University, Incheon, South Korea
\\
$^{52}$Institut de Physique Nucl\'eaire d'Orsay (IPNO), Universit\'e Paris-Sud, CNRS-IN2P3, Orsay, France
\\
$^{53}$Institute for Nuclear Research, Academy of Sciences, Moscow, Russia
\\
$^{54}$Institute for Subatomic Physics of Utrecht University, Utrecht, Netherlands
\\
$^{55}$Institute for Theoretical and Experimental Physics, Moscow, Russia
\\
$^{56}$Institute of Experimental Physics, Slovak Academy of Sciences, Ko\v{s}ice, Slovakia
\\
$^{57}$Institute of Physics, Academy of Sciences of the Czech Republic, Prague, Czech Republic
\\
$^{58}$Institute of Physics, Bhubaneswar, India
\\
$^{59}$Institute of Space Science (ISS), Bucharest, Romania
\\
$^{60}$Institut f\"{u}r Informatik, Johann Wolfgang Goethe-Universit\"{a}t Frankfurt, Frankfurt, Germany
\\
$^{61}$Institut f\"{u}r Kernphysik, Johann Wolfgang Goethe-Universit\"{a}t Frankfurt, Frankfurt, Germany
\\
$^{62}$Institut f\"{u}r Kernphysik, Westf\"{a}lische Wilhelms-Universit\"{a}t M\"{u}nster, M\"{u}nster, Germany
\\
$^{63}$Instituto de Ciencias Nucleares, Universidad Nacional Aut\'{o}noma de M\'{e}xico, Mexico City, Mexico
\\
$^{64}$Instituto de F\'{i}sica, Universidade Federal do Rio Grande do Sul (UFRGS), Porto Alegre, Brazil
\\
$^{65}$Instituto de F\'{\i}sica, Universidad Nacional Aut\'{o}noma de M\'{e}xico, Mexico City, Mexico
\\
$^{66}$Institut Pluridisciplinaire Hubert Curien (IPHC), Universit\'{e} de Strasbourg, CNRS-IN2P3, Strasbourg, France
\\
$^{67}$iThemba LABS, National Research Foundation, Somerset West, South Africa
\\
$^{68}$Joint Institute for Nuclear Research (JINR), Dubna, Russia
\\
$^{69}$Konkuk University, Seoul, South Korea
\\
$^{70}$Korea Institute of Science and Technology Information, Daejeon, South Korea
\\
$^{71}$KTO Karatay University, Konya, Turkey
\\
$^{72}$Laboratoire de Physique Corpusculaire (LPC), Clermont Universit\'{e}, Universit\'{e} Blaise Pascal, CNRS--IN2P3, Clermont-Ferrand, France
\\
$^{73}$Laboratoire de Physique Subatomique et de Cosmologie, Universit\'{e} Grenoble-Alpes, CNRS-IN2P3, Grenoble, France
\\
$^{74}$Laboratori Nazionali di Frascati, INFN, Frascati, Italy
\\
$^{75}$Laboratori Nazionali di Legnaro, INFN, Legnaro, Italy
\\
$^{76}$Lawrence Berkeley National Laboratory, Berkeley, California, United States
\\
$^{77}$Moscow Engineering Physics Institute, Moscow, Russia
\\
$^{78}$Nagasaki Institute of Applied Science, Nagasaki, Japan
\\
$^{79}$National Centre for Nuclear Studies, Warsaw, Poland
\\
$^{80}$National Institute for Physics and Nuclear Engineering, Bucharest, Romania
\\
$^{81}$National Institute of Science Education and Research, Bhubaneswar, India
\\
$^{82}$National Research Centre Kurchatov Institute, Moscow, Russia
\\
$^{83}$Niels Bohr Institute, University of Copenhagen, Copenhagen, Denmark
\\
$^{84}$Nikhef, Nationaal instituut voor subatomaire fysica, Amsterdam, Netherlands
\\
$^{85}$Nuclear Physics Group, STFC Daresbury Laboratory, Daresbury, United Kingdom
\\
$^{86}$Nuclear Physics Institute, Academy of Sciences of the Czech Republic, \v{R}e\v{z} u Prahy, Czech Republic
\\
$^{87}$Oak Ridge National Laboratory, Oak Ridge, Tennessee, United States
\\
$^{88}$Petersburg Nuclear Physics Institute, Gatchina, Russia
\\
$^{89}$Physics Department, Creighton University, Omaha, Nebraska, United States
\\
$^{90}$Physics Department, Panjab University, Chandigarh, India
\\
$^{91}$Physics Department, University of Athens, Athens, Greece
\\
$^{92}$Physics Department, University of Cape Town, Cape Town, South Africa
\\
$^{93}$Physics Department, University of Jammu, Jammu, India
\\
$^{94}$Physics Department, University of Rajasthan, Jaipur, India
\\
$^{95}$Physikalisches Institut, Eberhard Karls Universit\"{a}t T\"{u}bingen, T\"{u}bingen, Germany
\\
$^{96}$Physikalisches Institut, Ruprecht-Karls-Universit\"{a}t Heidelberg, Heidelberg, Germany
\\
$^{97}$Physik Department, Technische Universit\"{a}t M\"{u}nchen, Munich, Germany
\\
$^{98}$Purdue University, West Lafayette, Indiana, United States
\\
$^{99}$Pusan National University, Pusan, South Korea
\\
$^{100}$Research Division and ExtreMe Matter Institute EMMI, GSI Helmholtzzentrum f\"ur Schwerionenforschung, Darmstadt, Germany
\\
$^{101}$Rudjer Bo\v{s}kovi\'{c} Institute, Zagreb, Croatia
\\
$^{102}$Russian Federal Nuclear Center (VNIIEF), Sarov, Russia
\\
$^{103}$Saha Institute of Nuclear Physics, Kolkata, India
\\
$^{104}$School of Physics and Astronomy, University of Birmingham, Birmingham, United Kingdom
\\
$^{105}$Secci\'{o}n F\'{\i}sica, Departamento de Ciencias, Pontificia Universidad Cat\'{o}lica del Per\'{u}, Lima, Peru
\\
$^{106}$Sezione INFN, Bari, Italy
\\
$^{107}$Sezione INFN, Bologna, Italy
\\
$^{108}$Sezione INFN, Cagliari, Italy
\\
$^{109}$Sezione INFN, Catania, Italy
\\
$^{110}$Sezione INFN, Padova, Italy
\\
$^{111}$Sezione INFN, Rome, Italy
\\
$^{112}$Sezione INFN, Trieste, Italy
\\
$^{113}$Sezione INFN, Turin, Italy
\\
$^{114}$SSC IHEP of NRC Kurchatov institute, Protvino, Russia
\\
$^{115}$Stefan Meyer Institut f\"{u}r Subatomare Physik (SMI), Vienna, Austria
\\
$^{116}$SUBATECH, Ecole des Mines de Nantes, Universit\'{e} de Nantes, CNRS-IN2P3, Nantes, France
\\
$^{117}$Suranaree University of Technology, Nakhon Ratchasima, Thailand
\\
$^{118}$Technical University of Ko\v{s}ice, Ko\v{s}ice, Slovakia
\\
$^{119}$Technical University of Split FESB, Split, Croatia
\\
$^{120}$The Henryk Niewodniczanski Institute of Nuclear Physics, Polish Academy of Sciences, Cracow, Poland
\\
$^{121}$The University of Texas at Austin, Physics Department, Austin, Texas, United States
\\
$^{122}$Universidad Aut\'{o}noma de Sinaloa, Culiac\'{a}n, Mexico
\\
$^{123}$Universidade de S\~{a}o Paulo (USP), S\~{a}o Paulo, Brazil
\\
$^{124}$Universidade Estadual de Campinas (UNICAMP), Campinas, Brazil
\\
$^{125}$Universidade Federal do ABC, Santo Andre, Brazil
\\
$^{126}$University of Houston, Houston, Texas, United States
\\
$^{127}$University of Jyv\"{a}skyl\"{a}, Jyv\"{a}skyl\"{a}, Finland
\\
$^{128}$University of Liverpool, Liverpool, United Kingdom
\\
$^{129}$University of Tennessee, Knoxville, Tennessee, United States
\\
$^{130}$University of the Witwatersrand, Johannesburg, South Africa
\\
$^{131}$University of Tokyo, Tokyo, Japan
\\
$^{132}$University of Tsukuba, Tsukuba, Japan
\\
$^{133}$University of Zagreb, Zagreb, Croatia
\\
$^{134}$Universit\'{e} de Lyon, Universit\'{e} Lyon 1, CNRS/IN2P3, IPN-Lyon, Villeurbanne, Lyon, France
\\
$^{135}$Universit\`{a} di Brescia, Brescia, Italy
\\
$^{136}$V.~Fock Institute for Physics, St. Petersburg State University, St. Petersburg, Russia
\\
$^{137}$Variable Energy Cyclotron Centre, Kolkata, India
\\
$^{138}$Warsaw University of Technology, Warsaw, Poland
\\
$^{139}$Wayne State University, Detroit, Michigan, United States
\\
$^{140}$Wigner Research Centre for Physics, Hungarian Academy of Sciences, Budapest, Hungary
\\
$^{141}$Yale University, New Haven, Connecticut, United States
\\
$^{142}$Yonsei University, Seoul, South Korea
\\
$^{143}$Zentrum f\"{u}r Technologietransfer und Telekommunikation (ZTT), Fachhochschule Worms, Worms, Germany
\endgroup

  %%%%%%% done by webmaster team

\end{document}